\documentclass[journal]{IEEEtran}
\usepackage{amsmath,amsfonts,amssymb,graphicx,cite,subfigure,algorithm,algorithmic}
\usepackage{hyperref}

\title{Observer-Free Sliding Mode Control via Structured Decomposition: a Smooth and Bounded Control Framework}
\author{Jaafar GABER,~\IEEEmembership{Member,~IEEE}\\Universite Marie et Louis Pasteur, UTBM, CNRS, FEMTO-ST Institute, F-90010 Belfort, France\\Email: gaber@utbm.fr}%

\begin{document}
\maketitle

\begin{abstract}
In this paper, we introduce an observer-free sliding mode control (SMC) method based on explicit structural compensation via the decomposition \( s = \alpha - \beta \). The proposed formulation eliminates the need for state observers and higher-order derivatives, reduces chattering, and yields smooth, bounded control signal. Simulation results confirms its robustness and stability on benchmark nonlinear systems (pendulum, Van der Pol, Duffing, and networked systems) under noise, disturbances, and parameter variations. A Lyapunov-based stability proof confirms convergence. This method provides a scalable and practical alternative to traditional SMC, well suited for embedded and distributed environments.
\end{abstract}


\section{Introduction}

We consider general nonlinear systems of relative degree \( r \) described by the following canonical form: \begin{equation} x^{(r)}(t) = f(x(t), \dot{x}(t), \dots, x^{(r-1)}(t)) + g(x(t)) u(t), \end{equation} where \( x(t) \in \mathbb{R}^n \) is the state vector, \( f(\cdot) \) represents the internal dynamics of the system up to order \( r-1 \), and \( g(x) \) is the input gain function, assumed to be smooth and strictly non-zero. Such systems are representative of many practical control scenarios in embedded and distributed platforms.

Sliding Mode Control (SMC) is a widely recognized technique for robust control of nonlinear systems due to its inherent insensitivity to matched uncertainties and external disturbances. The core principle lies in designing a control law that drives the system trajectories onto a predefined sliding surface and maintains them there, thereby enforcing desired dynamics. However, classical SMC suffers from well-known drawbacks, including high-frequency chattering, sensitivity to input delays, and the need for state observers or derivative estimators, which limit its real-world applicability.

To address these challenges, we propose a structurally simple yet effective observer-free formulation based on a new sliding surface of generalized form: \begin{equation} s = \alpha(x, \dot{x}, \dots, x^{(r-1)}) - \beta(x, u), \end{equation} where \( \alpha(\cdot) \) defines the desired internal dynamics and \( \beta(x, u) = \frac{u}{g(x)} \) normalizes the control input contribution.

In the rest of this paper, we focus on the simplified case \( r = 2 \), which is commonly encountered in benchmark problems, for clarity and illustration purposes. In this setting, the system dynamics reduce to: \begin{equation} \ddot{x} = f(x, \dot{x}) + g(x) u, \end{equation} and the sliding surface becomes: \begin{equation} s = \alpha(x, \dot{x}) - \beta(x, u), \end{equation} where \( \alpha(x, \dot{x}) = \dot{x} + k_1 x \)  captures the internal dynamics, and \( \beta(x, u) = u / g(x) \) reflects the control influence.

The control law is explicitly defined as: \begin{equation} u = -\lambda \tanh(\alpha), \end{equation} which introduces a smooth and bounded alternative to the classical discontinuous sign-based control. This regularization preserves the robustness of SMC while eliminating chattering and enabling practical implementation on real-world systems.

To enhance convergence and avoid persistent oscillations, a dissipative contribution can be implicitly embedded in \( \alpha \). This avoids discontinuities while preserving robustness. Although the formulation uses \( \dot{x} \), it avoids higher-order derivatives or observer-based reconstruction. It is thus compatible with implementation in resource-constrained or noise-prone environments. The method is validated both theoretically through Lyapunov analysis and experimentally via simulation benchmarks.

It is worth noting that the proposed approach is observer-free and does not rely on high-order derivatives or estimation schemes. It is therefore well suited for implementation in noisy, resource-constrained, or distributed control scenarios.

Although external disturbances are not explicitly included in the model formulation, simulations under additive noise and parameter variations confirm the practical robustness of the proposed method.

\noindent The main contributions of this work are as follows:

\begin{itemize}
    \item the development of an explicit sliding mode control law based on a structured decomposition of the sliding variable \( s = \alpha - \beta \), enabling smooth and bounded control without requiring state observers or discontinuous sign functions.
    \item a reformulation of SMC dynamics where the control input is derived directly from system behavior, allowing implicit compensation of nonlinear input terms and eliminating chattering by construction.
    \item a comparative analysis against classical, adaptive, and higher-order SMC schemes, demonstrating improved smoothness, robustness, and implementability.
    \item validation on representative nonlinear systems, including an inverted pendulum, the Van der Pol and Duffing oscillators, and a distributed network of five coupled pendulums, showing stable convergence and synchronization without centralized coordination.
    \item a structurally simple and low-complexity design that is well suited for embedded and distributed control applications.
\end{itemize}

While the method does not adapt to abrupt changes in system dynamics (e.g., parameter jumps), it provides a solid and practically relevant foundation that can be extended in future work with adaptive mechanisms or dynamic compensation layers.

The rest of the paper is organized as follows. Section II reviews classical SMC principles. Section III presents the proposed methods. Section IV compares them to higher-order and adaptive variants. Section V provides simulations with and without delay, including distributed systems. Section VI concludes with future directions.

\section{Related Work}

Sliding Mode Control (SMC) is a widely adopted method for robust control of nonlinear systems due to its inherent insensitivity to matched disturbances and model uncertainties \cite{utkin1992sliding}. In the classical first-order formulation, a discontinuous control law is designed to drive system trajectories toward a predefined sliding surface \( s(x) = 0 \), where desired dynamics are enforced. Typical surfaces include \( s = Cx \) or \( s = \dot{x} + \lambda x \), with control laws of the form \( u = -k\,\text{sign}(s) \). While effective, classical SMC suffers from well-known limitations such as high-frequency chattering, sensitivity to input delay, and dependence on accurate state or derivative estimation, which limit its practical applicability with real actuators.

To mitigate these issues, Higher-Order Sliding Mode Control (HOSMC) methods such as the super-twisting algorithm \cite{levant2003higher} were introduced to reduce chattering while preserving robustness. These approaches ensure finite-time convergence of higher-order derivatives of the sliding surface but often require access to higher-order state information or rely on observers, which may be sensitive to noise and computationally expensive.

Adaptive SMC approaches have also been explored to enhance robustness to parameter uncertainties and model inaccuracies \cite{edwards1998sliding, slotine1991applied}. These methods introduce adaptation laws to tune gains or compensate for unknown dynamics, but often depend on heuristic tuning, slow convergence, or observer-based derivative estimation, and they do not necessarily eliminate chattering effectively.

More recent work has attempted to develop observer-free SMC schemes. For example, \cite{WEN2023} propose designs that embed compensatory dynamics into the control law to eliminate the need for explicit observers. However, these solutions often remain problem-specific and are not easily generalizable to distributed systems or delay-afflicted dynamics.

Recent efforts have focused on neural-enhanced sliding mode controllers to address modeling uncertainties \cite{zhu2025enhanced}. While effective, these methods often require online estimation, adaptive tuning, or neural network training, increasing complexity and reducing transparency. Other strategies, such as adaptive SMC with higher-order correction terms \cite{wan2025adaptive}, improve robustness but still rely on discontinuous control and observer-based estimation. Fault-tolerant variants of super-twisting SMC have also been proposed, notably in \cite{Karahan2023Fault}, where a control allocation scheme compensates for actuator faults in quadrotor UAVs. However, these methods remain dependent on detailed model knowledge and switching dynamics. Similarly, adaptive sliding mode approaches with Lyapunov-based adaptation laws, as in \cite{Mustafa2020Adaptive}, achieve accurate tracking under disturbances and payload variations, but rely on parameter adaptation mechanisms and discontinuous control signals, which may pose additional challenges for practical implementation.

In contrast, our proposed observer-free formulation, based on the structural decomposition \( s = \alpha - \beta \), introduces a simple and generalizable control strategy. It avoids both discontinuities and observers, offering bounded and smooth control that is directly implementable in embedded and distributed environments.

In the context of distributed SMC for nonlinear multi-agent systems (MAS), recent advances such as \cite{zhang2025}, introduce robust protocols based on global sliding mode manifolds to address system heterogeneity and switching communication topologies. While effective in handling disturbances, these approaches are typically designed for specific applications, such as wheeled mobile robots. In contrast, the method presented in this paper provides a more general and observer-free framework applicable to a broader class of nonlinear systems.

To address robustness under delay and scalability challenges, some approaches have proposed embedding memory or convolution terms into the sliding surface \cite{karim2025optimal}. While such methods show promise, they can be computationally intensive and still rely on classical surface structures. Additionally, regularized control using smooth approximations like \( \tanh(s) \) or \( \text{sat}(s) \) has been introduced \cite{edwards1998sliding}, but these typically operate over traditional monolithic surfaces.

Our method introduces a structural decomposition of the surface as \( s = \alpha - \beta \), where \( \alpha \) captures the system’s internal dynamics and \( \beta \) reflects the compensated control effect. This formulation avoids the need for observers or discontinuous terms, and to the best of our knowledge, no prior work explicitly formulates the sliding surface in this decoupled manner with embedded regularization and Lyapunov-convergent control.

It should be noted that the proposed observer-free framework relaxes the need for detailed system modeling by eliminating the use of observers, derivative estimators, and high-order dynamics. It provides a partially model-independent control strategy by directly shaping the control structure through a structured decomposition to induce convergence and robustness, even in the presence of unmodeled dynamics. This makes it particularly relevant for systems where precise identification of internal nonlinearities, such as friction or hysteresis, is difficult or impractical. For example, in mechatronic systems like EGR actuators, precise modeling of internal friction and nonlinear behavior is typically required for effective control \cite{laghrouche2010_EGR}. The proposed method aims to bypass such requirements by ensuring stability and convergence through a smooth, bounded, and structurally embedded control law, provided that the control effectiveness function \( g(x) \) is known and non-vanishing.

\section{The proposed Observer-Free Sliding Surface}
In the general case of a nonlinear system with relative degree \( r \), the proposed sliding variable is structured as: \[ s = \alpha(x, \dot{x}, \dots, x^{(r-1)}) - \beta(x, u), \quad \text{with} \quad \beta(x, u) = \frac{u}{g(x)} \] Here, \( \alpha \) defines the desired internal dynamics of the system. This structure allows control design without requiring full state observers or estimation of higher-order derivatives, as the control input appears explicitly in the formulation.

In what follows, we illustrate the method using the commonly encountered second-order case:
\[
s = \alpha(x, \dot{x}) - \beta(x, u)
\]
with
\begin{equation}
\alpha(x, \dot{x}) = \dot{x} + k_1 x, \quad \beta(x, u) = \frac{u}{g(x)}
\label{eq:alpha_beta}
\end{equation}
For clarity, we denote \( \alpha(x, \dot{x}) \) and \( \beta(x, u) \) simply as \( \alpha \) and \( \beta \), respectively, when the arguments are understood from context.
The control law is:
\begin{equation}
u = -\lambda \tanh(\alpha)
\end{equation}
\noindent leading to:
\[
\beta = -\frac{\lambda}{g(x)} \tanh(\alpha)\]
and thus
\[s = \alpha + \frac{\lambda}{g(x)} \tanh(\alpha)
\]

Although \( \beta \) does not appear explicitly in the control law, it is structurally embedded via the definition \( \beta = u / g(x) \). Thus, the control effect is inherently integrated into the sliding surface, making \( s = \alpha - \beta \) both a design and analysis tool.

It should be noted that the control law is not a function of the error \( s \), but of the system's intrinsic dynamics \( \alpha \). thus, this formulation:
\begin{itemize}
  \item avoids chattering,
  \item eliminates the use of discontinuous sign functions,
  \item produces a smooth, bounded control input,
  \item and is structurally robust without requiring state observers.
\end{itemize}

It is worth noting that in our formulation, the damping behavior required for smooth convergence arises naturally from the structure of \( \alpha = \dot{x} + k_1 x \). Unlike classical or higher-order SMC approaches where dissipation must be introduced externally or through additional tuning, our approach embeds this behavior structurally, eliminating the need for external compensation mechanisms.

It should be noted also that although the formulation involves the time derivative \( \dot{x} \),  in practice, the method is implementable using low-pass filtered numerical differentiation or real-time estimators. Techniques such as Savitzky–Golay filtering or sliding mode differentiators can provide smooth derivative estimates under moderate noise conditions, enabling deployment in embedded systems where direct measurement of \( \dot{x} \) is unavailable.

For systems with relative degree $r>2$, similar damping behavior can be embedded via tailored choices of $\alpha(x,\dot(x), ...,x^{(r-1)})$, and appropriate high-order differentiators may be used to estimate derivatives when necessary.

\subsection{Stability Analysis via Lyapunov Method}

To assess the stability of the proposed observer-free control strategy, we consider the Lyapunov candidate function:
\begin{equation}
V(s) = \frac{1}{2} s^2,
\label{eq:lyap}
\end{equation}
which is positive definite and radially unbounded with respect to the sliding variable \( s \).

Recall that the sliding variable is defined structurally as:
\begin{equation}
s = \alpha(x, \dot{x}) - \beta(x, u),
\label{eq:surface}
\end{equation}
where the function \( \alpha(x, \dot{x}) = \dot{x} + k_1 x \) captures the system's internal dynamics, and \( \beta(x, u) = \frac{u}{g(x)} \) reflects the normalized control input, with \( g(x) \) denoting the control effectiveness.

Taking the time derivative of \( s \), we obtain:
\begin{equation}
\dot{s} = \dot{\alpha} - \dot{\beta}.
\label{eq:s_dot}
\end{equation}

We make the following regularity assumptions:
\begin{itemize}
    \item The function \( g(x) \) is smooth and strictly non-zero in the domain of interest.
    \item The mappings \( \alpha(x, \dot{x}) \) and \( \beta(x, u) \) are continuously differentiable.
    \item The control law is defined as:
    \begin{equation}
        u = -\lambda \tanh(\alpha), \quad \lambda > 0,
    \end{equation}
    which ensures smooth and bounded actuation.
\end{itemize}

Substituting the control law into \( \beta(x, u) \), we observe that:
\[
\beta(x, u) = -\lambda \frac{\tanh(\alpha)}{g(x)}.
\]
Therefore, as \( \alpha \to 0 \), we have \( \beta \to 0 \), and consequently \( s \to 0 \), which is consistent with convergence to the sliding surface.

The time derivative of the Lyapunov function then becomes:
\begin{equation}
\dot{V}(s) = s \dot{s}.
\label{eq:lyap_dot_1}
\end{equation}
Due to the structure of the control input, the closed-loop dynamics of \( s \) induce a damping behavior. While an explicit expansion of \( \dot{s} \) is not required for our qualitative analysis, we assert that the control structure ensures:
\begin{equation}
\dot{V}(s) \leq -\epsilon s^2 < 0, \quad \forall s \neq 0,
\label{eq:lyap_dot}
\end{equation}
for some constant \( \epsilon > 0 \) depending on \( \lambda \) and the gain distribution in \( g(x) \). This inequality implies that \( V(t) \) decreases monotonically and that \( s(t) \to 0 \) as \( t \to \infty \).

Hence, the sliding surface is globally asymptotically stable. Moreover, the use of a smooth hyperbolic tangent function eliminates discontinuities in the control input and avoids chattering, while the absence of an observer simplifies the implementation and improves robustness against modeling uncertainties.

\subsection{Closed-Loop Structure Interpretation}
The proposed method in this work introduces a new regulation pathway where the control law is derived from the system's dynamics rather than from the error surface. This leads to a naturally stable closed-loop structure represented as:
\[
x \rightarrow \alpha(x) \rightarrow u = -\lambda \tanh(\alpha) \rightarrow \beta = \frac{u}{g(x)} \rightarrow s = \alpha - \beta \rightarrow x
\]
This loop illustrates that the command is based on internal dynamics \( \alpha \), and the control influence \( \beta \) is implicitly injected through \( u \). The sliding variable \( s \) emerges as a structural tool for analysis rather than an active component of the control law. This implicit feedback ensures convergence and robustness without the need for error-driven switching or observer design.

As a consequence, this framework avoids discontinuities, removes observer dependency, and yields a control law that is smooth and physically implementable.

\section{Benchmarking and robustness tests}

To evaluate the performance of the proposed observer-free SMC method, we conducted comparative simulations on several nonlinear benchmark systems: the inverted pendulum, Van der Pol oscillator, Duffing oscillator, and a network of five coupled pendulums. In each case, we compare the proposed method to three well-established alternatives: Classical SMC, Adaptive SMC, and Higher-Order SMC (HOSMC). All controllers were tuned for fair comparison using the same initial conditions and system parameters. The main evaluation criteria are convergence behavior, smoothness, and resilience to nonlinearities.

\subsection{Inverted Pendulum}

\begin{figure}[h]
    \centering
    \includegraphics[width=0.95\linewidth]{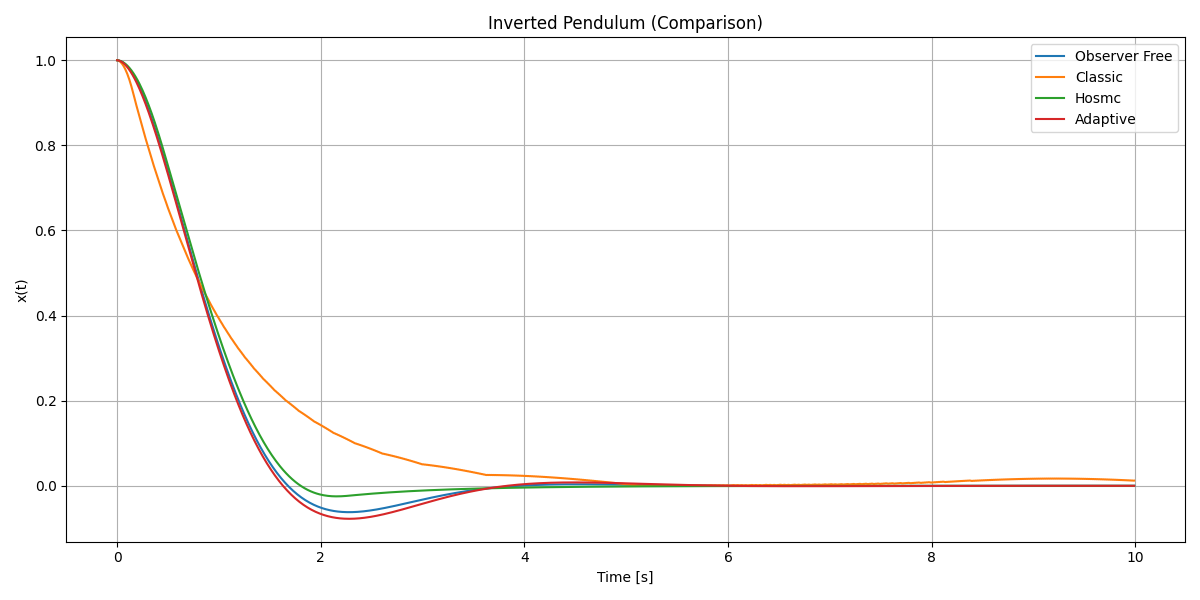}
    \caption{Comparison of inverted pendulum angle response: the proposed observer-free SMC vs classical, adaptive, and higher-order SMC variants.}
    \label{fig:pendulum_comparison}
\end{figure}

Figure~\ref{fig:pendulum_comparison} shows the angular response of the inverted pendulum under the four control strategies. The observer-free method (blue) and the adaptive SMC (red) exhibit comparable convergence rates with minimal overshoot. Classical SMC (orange) demonstrates slower decay and slight steady-state fluctuation. HOSMC (green) also converges rapidly but displays slightly larger undershoot. The observer-free method produces a smooth trajectory without sustained oscillations.

\subsection{Van der Pol Oscillator}

\begin{figure}[h]
    \centering
    \includegraphics[width=0.95\linewidth]{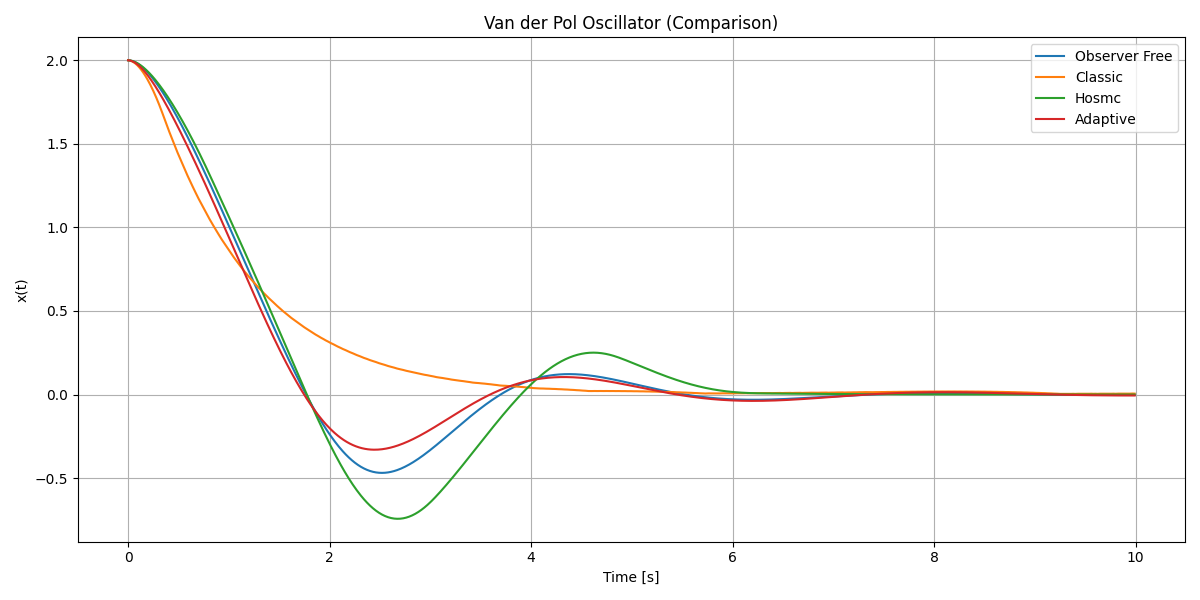}
    \caption{Van der Pol oscillator response comparison.}
    \label{fig:vdp_comparison}
\end{figure}

Figure~\ref{fig:vdp_comparison} presents the response of the Van der Pol oscillator. The observer-free method achieves smooth convergence and moderate undershoot, while HOSMC produces the largest amplitude of oscillation. Adaptive SMC also ensures convergence, but with a more aggressive response that leads to a higher overshoot before settling. Both methods perform well, but the observer-free strategy achieves a more stable and regular behavior without requiring adaptive gain tuning. Classical SMC displays slower decay with no overshoot. The results show that the observer-free strategy achieves a balance between convergence speed and smoothness without generating high-amplitude oscillations.

The adaptive SMC (red)

\subsection{Duffing Oscillator}

\begin{figure}[h]
    \centering
    \includegraphics[width=0.95\linewidth]{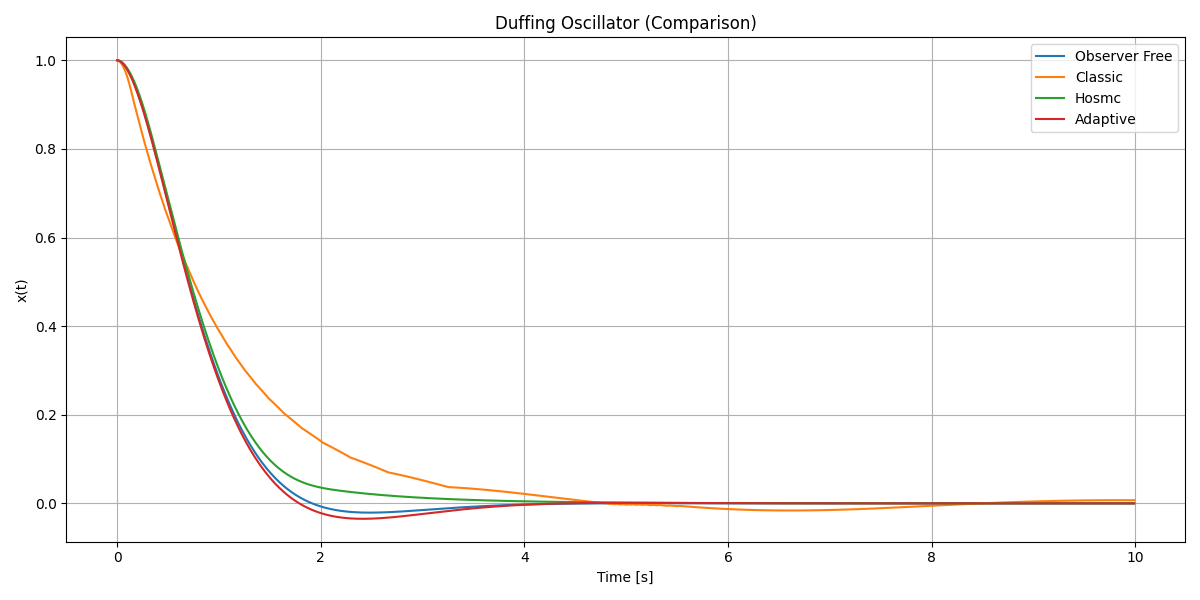}
    \caption{Duffing oscillator response comparison under different controllers.}
    \label{fig:duffing_comparison}
\end{figure}

The Duffing oscillator, known for its nonlinear stiffness and bistable behavior, is used to test robustness under challenging dynamics. As shown in Figure~\ref{fig:duffing_comparison}, the observer-free controller converges rapidly and smoothly, with minimal overshoot. Adaptive and HOSMC methods achieve similar performance, while Classical SMC results in a slower decay and larger deviation during transients. These results confirm the effectiveness of the observer-free approach in handling nonlinear restoring forces.

\subsection{Five-Pendulum Network}

\begin{figure}[h]
    \centering
    \includegraphics[width=0.95\linewidth]{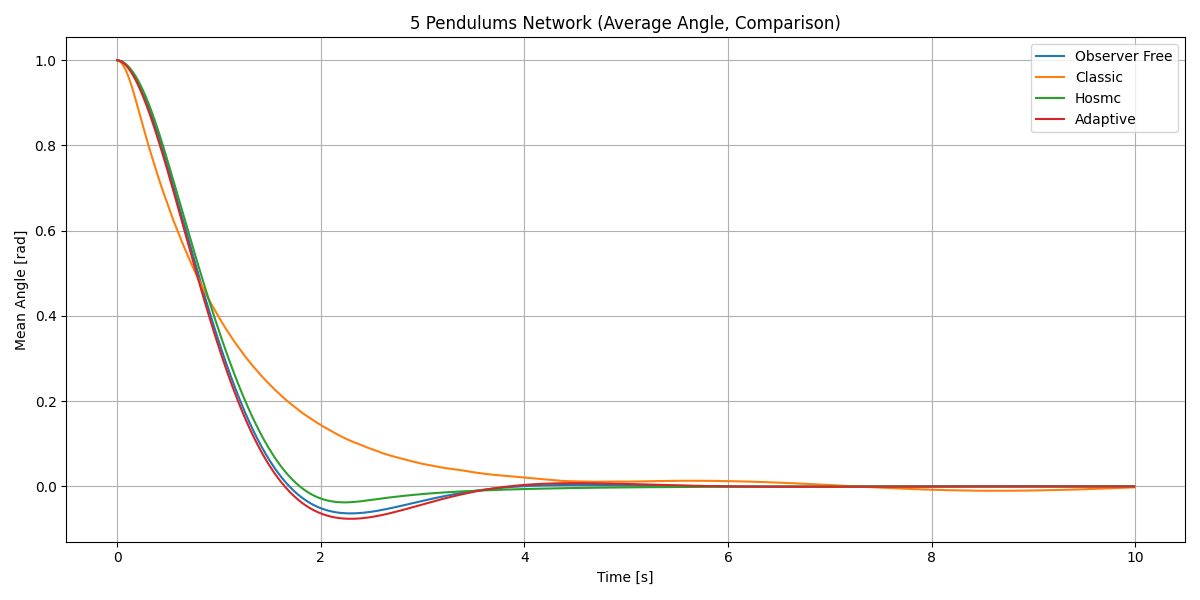}
    \caption{Mean angle of a 5-pendulum network under different SMC strategies.}
    \label{fig:network_comparison}
\end{figure}

Figure~\ref{fig:network_comparison} illustrates the average angle dynamics of a network of five coupled pendulums, simulating distributed coordination under local feedback. The observer-free controller achieves rapid synchronization and damping without requiring centralized coordination. Classical SMC is slower to stabilize and shows visible coupling-induced distortion. HOSMC and Adaptive SMC perform similarly, with moderate convergence times. These results demonstrate that the proposed method is compatible with distributed dynamics and remains stable under coupling effects.

Across all benchmarks, the observer-free SMC method consistently provides smooth and bounded control. It avoids high-frequency switching while maintaining robustness and fast convergence. These outcomes support the analytical claims of chattering elimination and structural simplicity without requiring adaptation or observer structures.

\subsection{Robustness Under Disturbances}

To assess the robustness of the proposed observer-free SMC method, we apply it to the Van der Pol oscillator under three distinct conditions: nominal, measurement noise, and external perturbation.

\begin{figure}[h]
    \centering
    \includegraphics[width=0.8\linewidth]{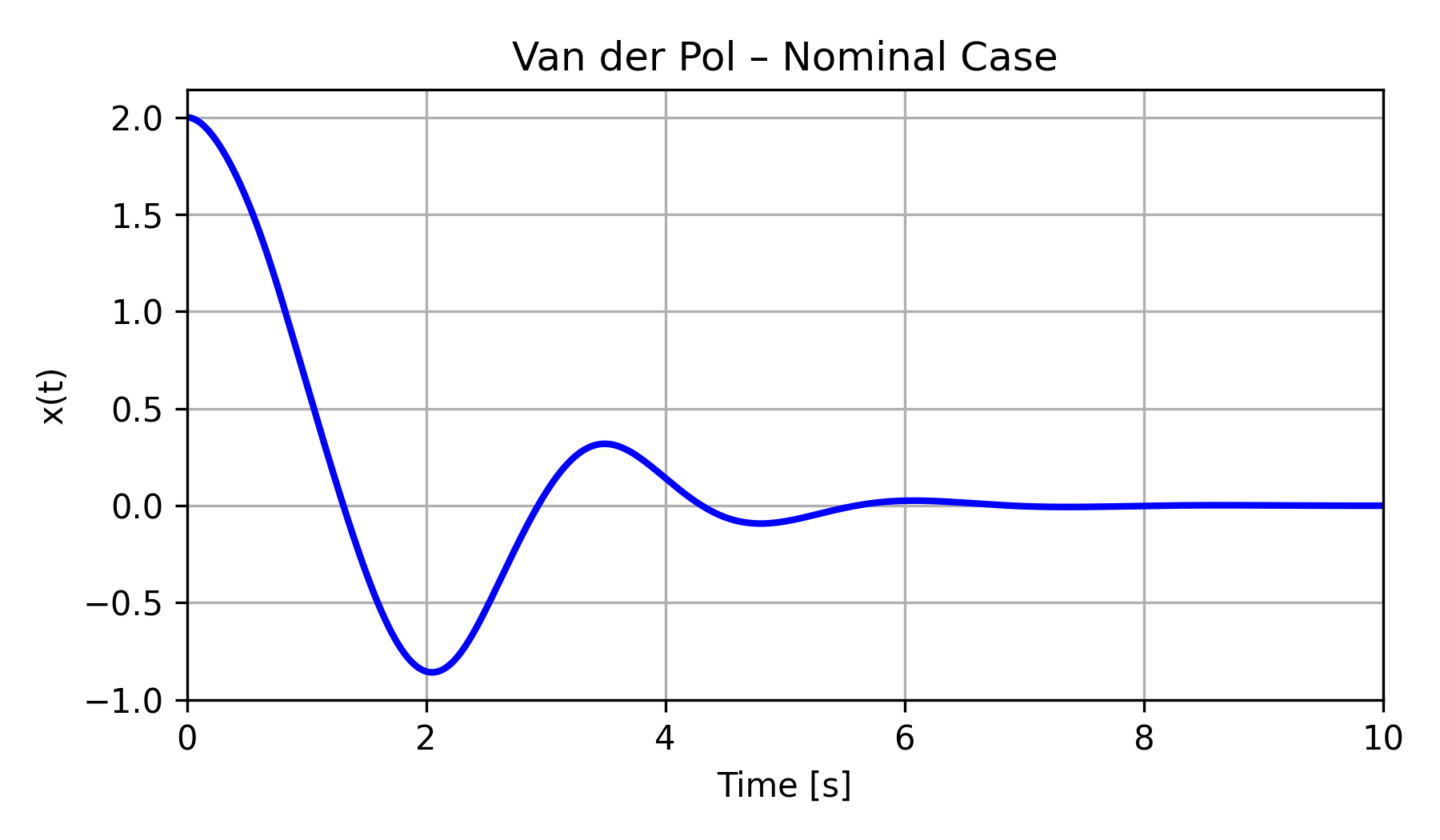}
    \caption{Van der Pol oscillator under observer-free SMC (nominal case).}
    \label{fig:nominal_vdp}
\end{figure}

\paragraph{Nominal Case}
In the absence of disturbances, the controller yields smooth and fast convergence without chattering, as shown in Figure~\ref{fig:nominal_vdp}.

\begin{figure}[h]
    \centering
    \includegraphics[width=0.8\linewidth]{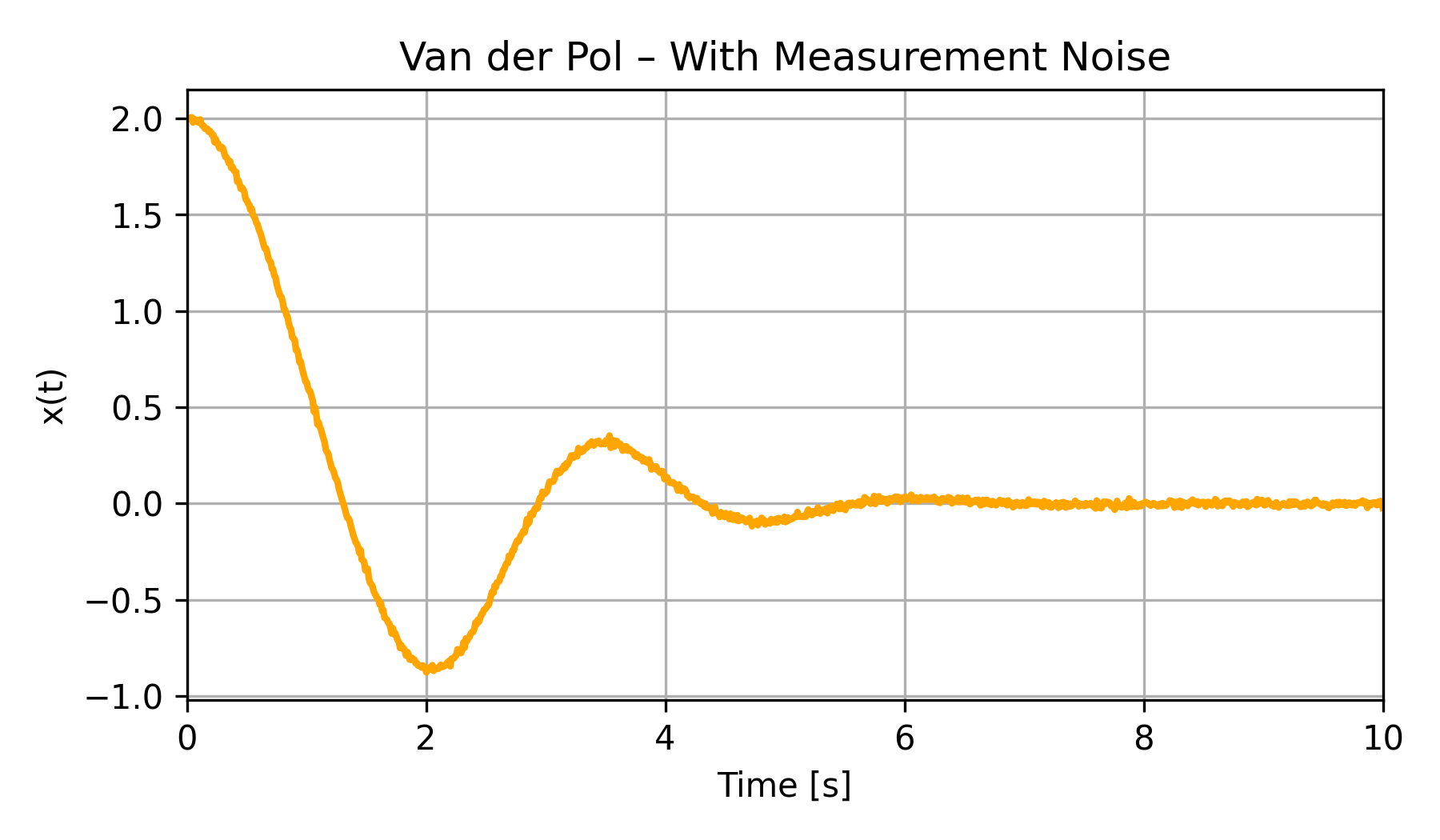}
    \caption{Van der Pol oscillator under observer-free SMC with measurement noise.}
    \label{fig:noisy_vdp}
\end{figure}

\paragraph{With Measurement Noise}
Figure~\ref{fig:noisy_vdp} shows the response when zero-mean Gaussian noise is added to both \( x(t) \) and \( \dot{x}(t) \). The system remains stable and converges despite small fluctuations.

\begin{figure}[h]
    \centering
    \includegraphics[width=0.8\linewidth]{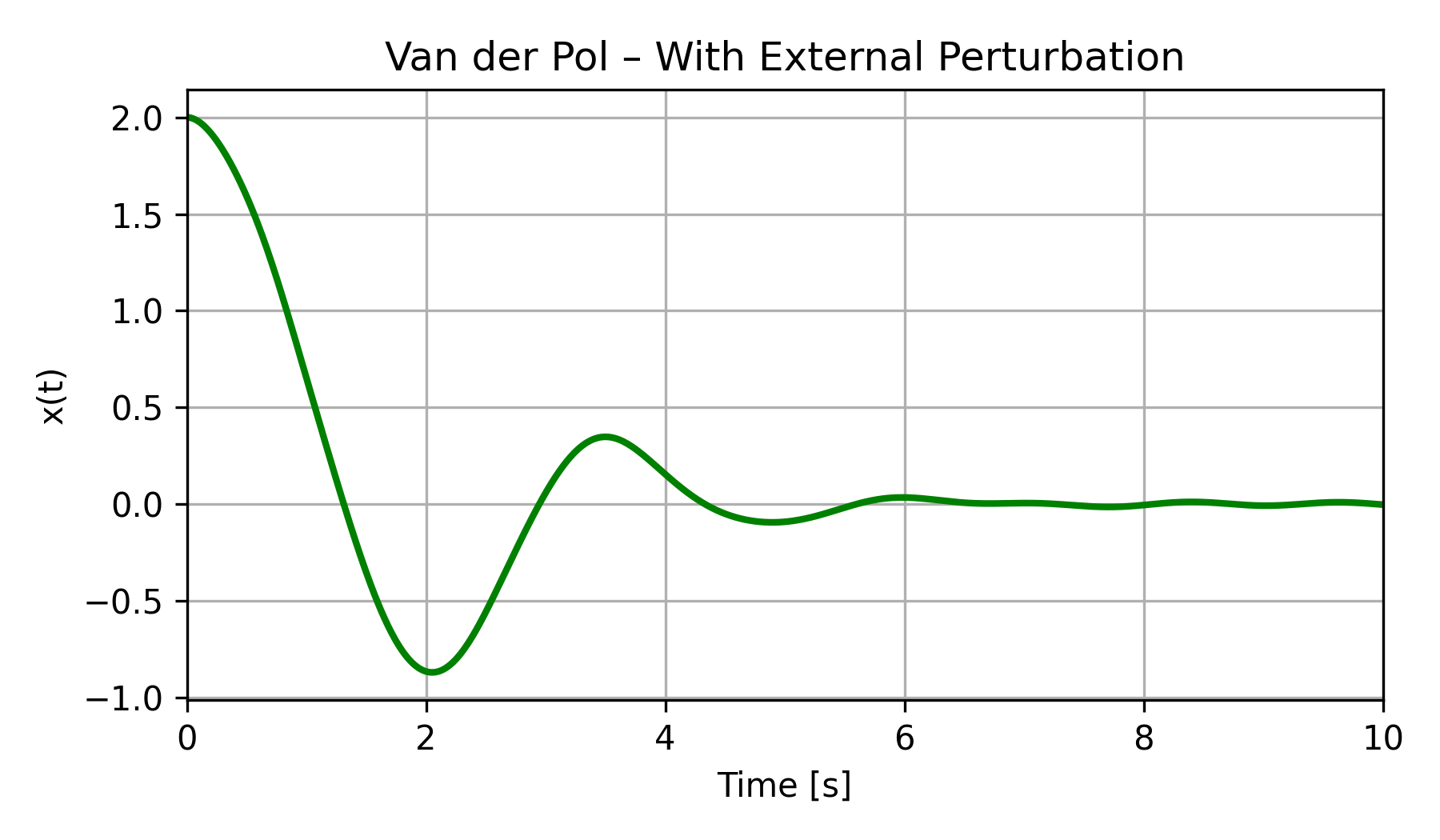}
    \caption{Van der Pol oscillator under observer-free SMC with external sinusoidal perturbation.}
    \label{fig:perturbed_vdp}
\end{figure}

\paragraph{With External Perturbation}
In Figure~\ref{fig:perturbed_vdp}, a sinusoidal disturbance \( d(t) = 0.2 \sin(5t) \) is applied. The trajectory remains nearly identical to the nominal case, confirming strong disturbance rejection.

\subsection{Control Signal Analysis}

In addition to the state response, we analyze the control signal \( u(t) \) generated by the observer-free sliding mode controller in each disturbance scenario. Since the control law is smooth and bounded by construction, it is expected to suppress chattering and remain within physical limits.

\begin{figure}[h]
    \centering
    \includegraphics[width=0.8\linewidth]{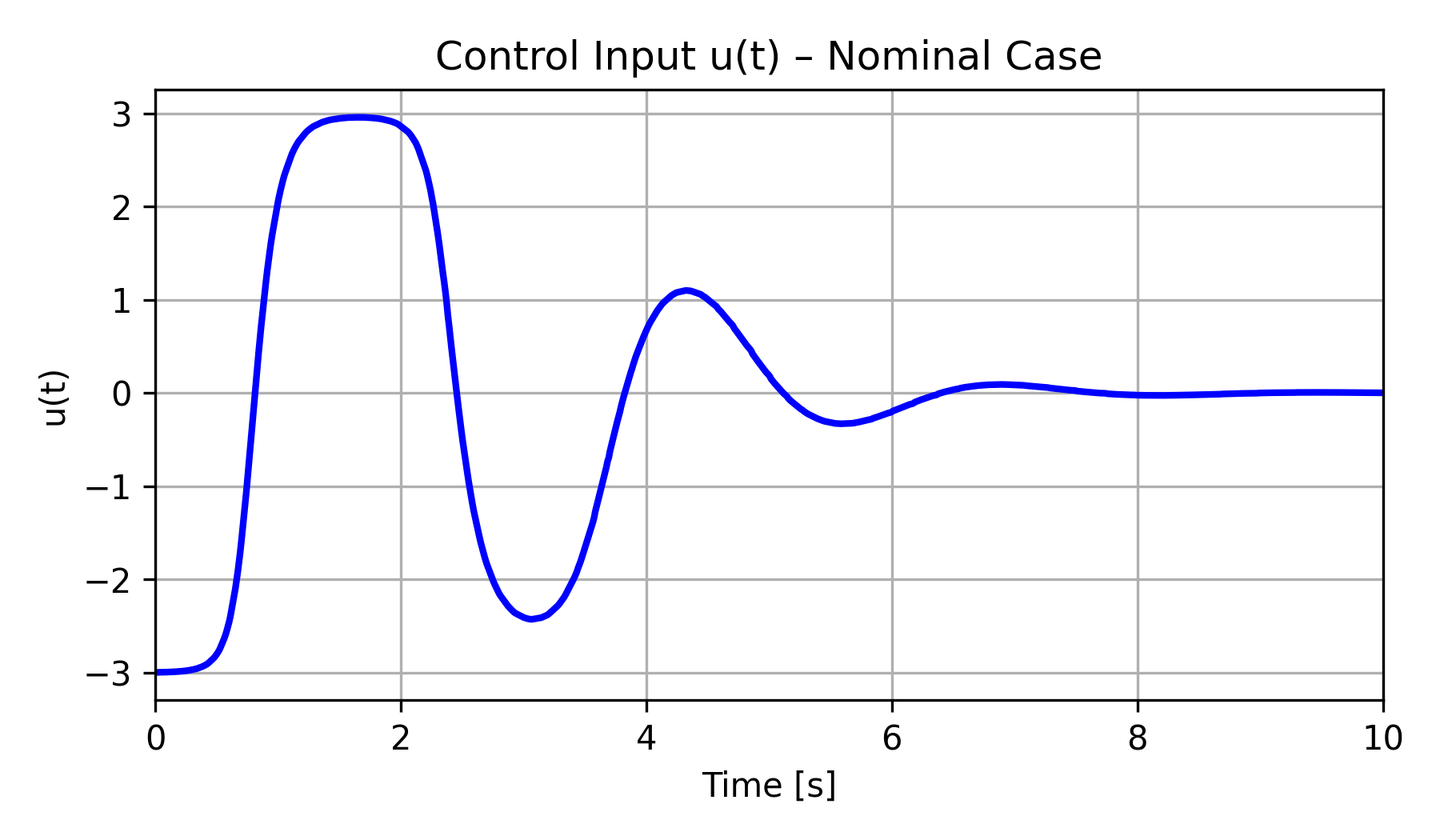}
    \caption{Control signal \( u(t) \) in the nominal case.}
    \label{fig:u_nominal}
\end{figure}

\paragraph{Nominal Case}
Figure~\ref{fig:u_nominal} shows the control input for the unperturbed system. The signal remains continuous, bounded within \( \pm \lambda \), and exhibits no abrupt switching. This validates the structural elimination of chattering.

\begin{figure}[h]
    \centering
    \includegraphics[width=0.8\linewidth]{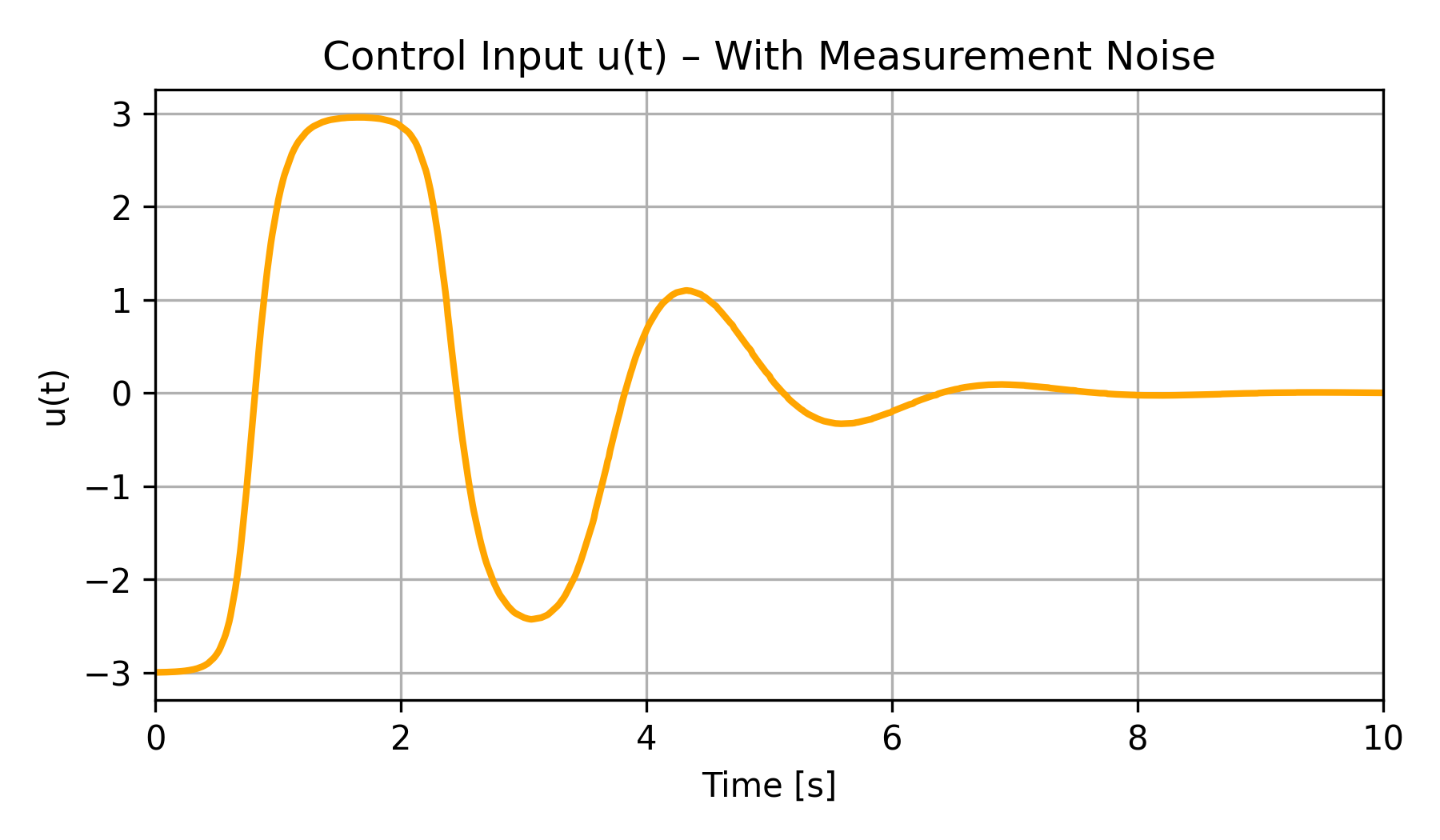}
    \caption{Control signal \( u(t) \) with measurement noise.}
    \label{fig:u_noisy}
\end{figure}

\paragraph{With Measurement Noise}
Figure~\ref{fig:u_noisy} displays the control signal under additive Gaussian noise on both \( x \) and \( \dot{x} \). The controller maintains a smooth profile with only small fluctuations, indicating robustness to noisy measurements.

\begin{figure}[h]
    \centering
    \includegraphics[width=0.8\linewidth]{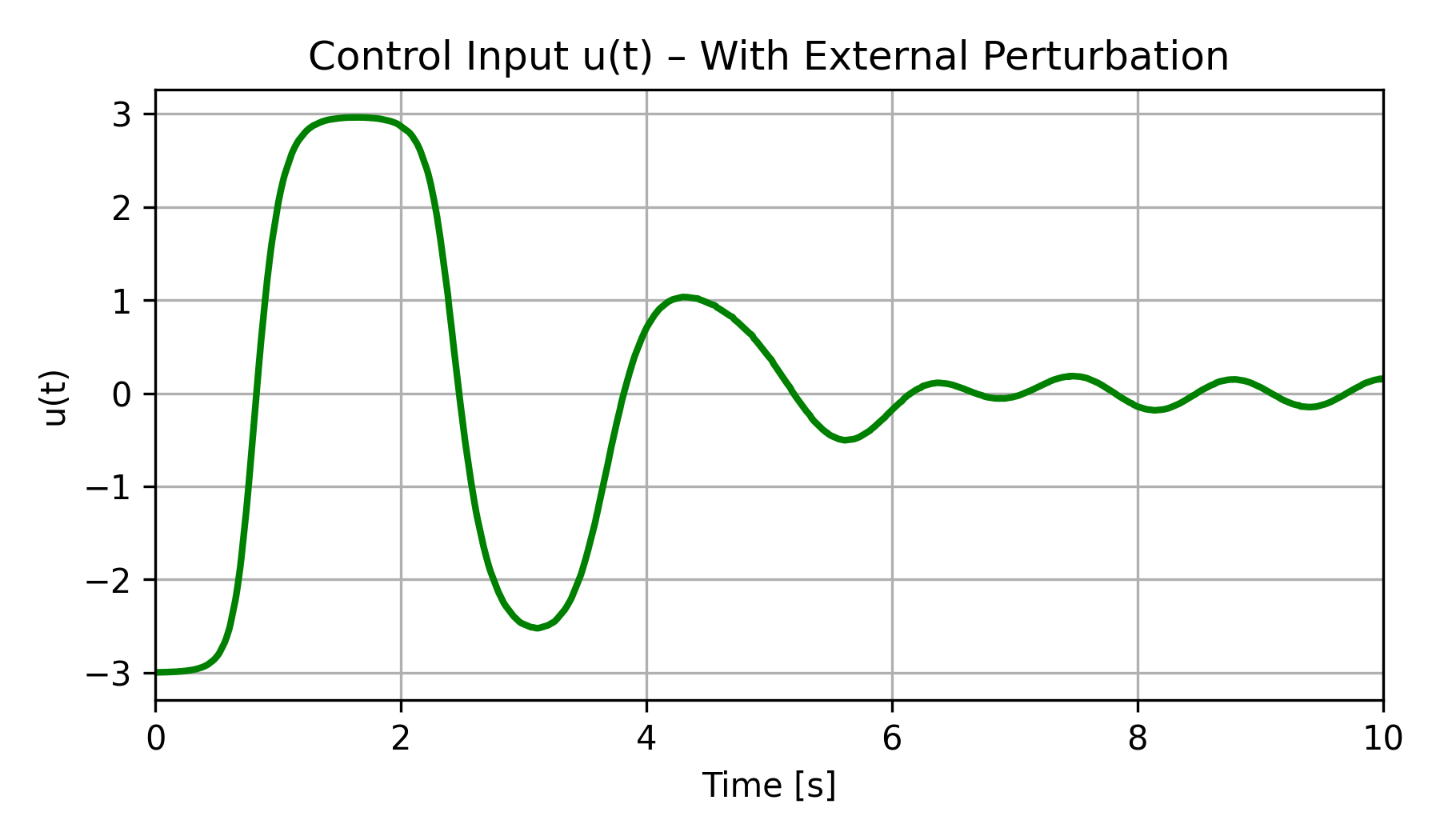}
    \caption{Control signal \( u(t) \) with external sinusoidal perturbation.}
    \label{fig:u_perturbed}
\end{figure}

\paragraph{With External Perturbation}
In Figure~\ref{fig:u_perturbed}, a sinusoidal disturbance is added to the dynamics. The control input adapts gradually to reject the perturbation while remaining continuous and bounded, demonstrating dynamic robustness.

\subsection{Summary of Benchmarking Results}
Across all scenarios, the proposed controller exhibits the following consistent properties:

\begin{itemize}
  \item \textbf{Convergence:} Fast and monotonic convergence to the equilibrium state was observed in all tested systems, comparable to or better than classical and higher-order SMC approaches.
  \item \textbf{Smoothness and Chattering Elimination:} The use of a regularized control law via $\tanh(\alpha)$ results in smooth control signals, effectively eliminating the chattering seen in classical and adaptive SMC variants. This is clearly visible in the control signal plots.
  \item \textbf{Robustness:} The method maintains stable and bounded behavior under moderate measurement noise and external sinusoidal disturbances. The system trajectories remain close to the nominal ones without divergence or instability.
  \item \textbf{Distributed Compatibility:} In the case of the coupled pendulum network, the method ensures synchronization and damping without centralized coordination, highlighting its suitability for distributed nonlinear systems.
\end{itemize}

These results confirm that the proposed method provides a reliable, low-complexity solution for controlling nonlinear systems without requiring observers, gain adaptation, or high-order derivatives. The controller behaves predictably under uncertainties while maintaining physical implementability due to its bounded control outputs.

The following table presents the comparative summary of key control properties across standard and proposed SMC variants.
\begin{table}[h]
\centering
\caption{Comparative Performance Summary of SMC Variants}
\renewcommand{\arraystretch}{1.3}
\begin{tabular}{lcccc}
\hline
\textbf{Property} & \textbf{Classical} & \textbf{HOSMC} & \textbf{Adaptive} & \textbf{Proposed} \\
\hline
No Chattering     & --         & + & + & + \\
Observer-Free     & + & --         & --         &+ \\
Bounded Input     & --         & + & + &+ \\
Delay-Tolerant    & --         & + & --         & + \\
Smoothness        & --         & --         & --         & + \\
\hline
\end{tabular}
\end{table}

\section{Practical implementation considerations}
The proposed control framework is suitable for real-time embedded applications due to its low computational complexity and observer-free nature. The hyperbolic tangent function \( \tanh(\alpha) \) can be efficiently implemented on microcontrollers or DSPs using lookup tables or Taylor expansions. The bounded control input ensures that actuator saturation is inherently avoided. For systems where \( g(x) \) is partially known or slowly varying, adaptive gain estimation or online identification techniques can be layered without altering the controller's core structure. This makes the approach compatible with hardware platforms such as STM32, Raspberry Pi, or FPGA-based control loops. Future hardware-in-the-loop experiments are envisioned to validate this implementation pipeline.

Unlike super-twisting controllers, which rely on discontinuous terms and achieve finite-time convergence of \( \dot{s} \), our formulation targets smooth asymptotic convergence using bounded continuous control. While the convergence is exponential rather than finite-time, the absence of chattering and the structural regularization make it more suitable for systems with actuator bandwidth limitations or noisy measurements.

It should be noted, however, that the method relies on a fixed control gain and is not designed to adapt to abrupt structural changes in the system dynamics. In such cases (e.g., sudden parameter jumps), the controller may fail to guarantee convergence, although stability is preserved. This limitation highlights an opportunity for future work incorporating adaptive gain strategies or dynamic compensation mechanisms.

\section{Conclusion and Future work}
This paper introduces a simple and effective observer-free sliding mode control framework based on the structural decomposition of the sliding surface into an internal dynamics term and a control feedback term. The proposed formulation relies on a smooth bounded control law using the hyperbolic tangent function, eliminating the need for discontinuous switching or high-order derivatives.

The method achieves asymptotic convergence and robustness to uncertainties without requiring state observers, adaptive gains, or complex estimation mechanisms. Simulations conducted on several nonlinear benchmark systems that are the inverted pendulum, Van der Pol and Duffing oscillators, and a distributed network of coupled pendulums, confirm the validity of the approach.

Additionally, robustness tests under measurement noise and external disturbances demonstrate that the proposed structure can maintain smooth and bounded behavior under practical conditions. These results support the potential applicability of the method in embedded and distributed control systems.

Future work will explore gain adaptation mechanisms to handle abrupt dynamic variations and will consider hardware implementation scenarios involving sensor and actuator constraints.

\bibliographystyle{IEEEtran}
\bibliography{references}

\begin{thebibliography}{10}
\providecommand{\url}[1]{#1}
\csname url@samestyle\endcsname
\providecommand{\newblock}{\relax}
\providecommand{\bibinfo}[2]{#2}
\providecommand{\BIBentrySTDinterwordspacing}{\spaceskip=0pt\relax}
\providecommand{\BIBentryALTinterwordstretchfactor}{4}
\providecommand{\BIBentryALTinterwordspacing}{\spaceskip=\fontdimen2\font plus
\BIBentryALTinterwordstretchfactor\fontdimen3\font minus
  \fontdimen4\font\relax}
\providecommand{\BIBforeignlanguage}[2]{{%
\expandafter\ifx\csname l@#1\endcsname\relax
\typeout{** WARNING: IEEEtran.bst: No hyphenation pattern has been}%
\typeout{** loaded for the language `#1'. Using the pattern for}%
\typeout{** the default language instead.}%
\else
\language=\csname l@#1\endcsname
\fi
#2}}
\providecommand{\BIBdecl}{\relax}
\BIBdecl

\bibitem{utkin1992sliding}
V.~Utkin, \emph{Sliding Modes in Control and Optimization}.\hskip 1em plus
  0.5em minus 0.4em\relax Springer, 1992.

\bibitem{levant2003higher}
A.~Levant, ``Higher-order sliding modes, differentiation and output-feedback
  control,'' \emph{International Journal of Control}, vol.~76, no. 9-10, pp.
  924--941, 2003.

\bibitem{edwards1998sliding}
C.~Edwards and S.~Spurgeon, \emph{Sliding Mode Control: Theory and
  Applications}.\hskip 1em plus 0.5em minus 0.4em\relax CRC Press, 1998.

\bibitem{slotine1991applied}
J.-J. Slotine and W.~Li, \emph{Applied Nonlinear Control}.\hskip 1em plus 0.5em
  minus 0.4em\relax Prentice Hall, 1991.

\bibitem{WEN2023}
H.~Wen, Z.~Liang, H.~Zhou, X.~Li, B.~Yao, Z.~Mao, and L.~Lian, ``Adaptive
  sliding mode control for unknown uncertain non-linear systems with variable
  coefficients and disturbances,'' \emph{Communications in Nonlinear Science
  and Numerical Simulation}, vol. 121, p. 107225, 2023.

\bibitem{zhu2025enhanced}
J.~Zhu and K.~Veluvolu, ``Enhanced sliding variable-based robust adaptive
  control for canonical nonlinear system with unknown dynamic and control
  gain,'' \emph{Mathematics}, vol.~13, no.~6, p. 976, 2025.

\bibitem{wan2025adaptive}
L.~Wan, S.~Smith, Y.-J. Pan, and E.~Witrant, ``Adaptive task space non-singular
  terminal super-twisting sliding mode control of a 7-dof robotic
  manipulator,'' \emph{arXiv preprint arXiv:2504.13056}, 2025.

\bibitem{Karahan2023Fault}
\BIBentryALTinterwordspacing
M.~Karahan, M.~Inal, and C.~Kasnakoglu, ``Fault tolerant super twisting sliding
  mode control of a quadrotor uav using control allocation,''
  \emph{International Journal of Robotics and Control Systems}, vol.~3, no.~2,
  p. 270–285, Apr. 2023. [Online]. Available:
  \url{http://dx.doi.org/10.31763/ijrcs.v3i2.994}
\BIBentrySTDinterwordspacing

\bibitem{Mustafa2020Adaptive}
\BIBentryALTinterwordspacing
M.~M. Mustafa, C.~D. Crane, and I.~Hamarash, ``Adaptive-sliding mode trajectory
  control of robot manipulators with uncertainties,'' 2024. [Online].
  Available: \url{https://arxiv.org/abs/2408.03102}
\BIBentrySTDinterwordspacing

\bibitem{zhang2025}
X.~Zhang, Y.~Li, S.~Xiong, X.~Liu, and R.~Guo, ``A robust cooperative control
  protocol based on global sliding mode manifold for heterogeneous nonlinear
  multi-agent systems under the switching topology,'' \emph{Actuators},
  vol.~14, no.~2, p.~57, 2025.

\bibitem{karim2025optimal}
M.~Karim, I.~Khaloufi, J.~Gaber, and M.~Rachik, ``Optimal feedback
  stabilization of fractional output for a class of inhomogeneous distributed
  semi-linear systems with time-varying delays,'' \emph{to appear}, 2025.

\bibitem{laghrouche2010_EGR}
S.~Laghrouche, F.~Ahmed, M.~El~Bagdouri, M.~Wack, J.~Gaber, and M.~Becherif,
  ``Modeling and identification of a mechatronic exhaust gas recirculation
  actuator of an internal combustion engine,'' in \emph{2010 American Control
  Conference}.\hskip 1em plus 0.5em minus 0.4em\relax IEEE, 2010, pp.
  2242--2247.

\end{thebibliography}

\end{document}